\documentclass[aps,prb,twocolumn, superscriptaddress,showpacs]{revtex4}
\usepackage{color}
\usepackage{booktabs}
\usepackage{graphicx}
\usepackage{epsfig}
\usepackage{amssymb}
\usepackage{bm}
\usepackage{graphpap}
\usepackage[normalem]{ulem}
\usepackage[dvipdfmx]{hyperref}
\usepackage[all]{hypcap}
\usepackage{url}
\hypersetup{colorlinks=true,citecolor=blue,linkcolor=red}

\begin{document}
\newdimen\heavyrulewidth
\newdimen\lightrulewidth
\newdimen\cmidrulewidth
\newdimen\belowrulesep
\newdimen\belowbottomsep
\newdimen\aboverulesep
\newdimen\abovetopsep
\newdimen\defaultaddspace
\heavyrulewidth=.08em
\lightrulewidth=.05em
\belowrulesep=.65ex
\belowbottomsep=.0pt
\aboverulesep=.4ex
\abovetopsep=.0pt
\defaultaddspace=.5em

\title{ High Luminescence in Small Si/SiO$_2$ Nanocrystals: A Theoretical Study }

\author{\firstname{Roberto} \surname{Guerra}}
\affiliation{Centro S3, CNR-Istituto di Nanoscienze, via Campi 213/A I-41100 Modena Italy.}
\affiliation{Dipartimento di Fisica, Universit\`a di Modena e Reggio Emilia, via Campi 213/A I-41100 Modena Italy.}
\author{\firstname{Stefano} \surname{Ossicini}}
\affiliation{Centro S3, CNR-Istituto di Nanoscienze, via Campi 213/A I-41100 Modena Italy.}
\affiliation{Dipartimento di Scienze e Metodi dell'Ingegneria, Universit\`a di Modena e Reggio Emilia, via Amendola 2 Pad. Morselli I-42100 Reggio Emilia Italy.}

\begin{abstract}
In recent years many experiments have demonstrated the possibility to achieve efficient photoluminescence from Si/SiO$_2$ nanocrystals. While it is widely known that only a minor portions of the nanocrystals in the samples contribute to the observed photoluminescence, the high complexity of the Si/SiO$_2$ interface and the dramatic sensitivity to the fabrication conditions make the identification of the most active structures at the experimental level not a trivial task. Focusing on this aspect we have addressed the problem theoretically, by calculating the radiative recombination rates for different classes of Si-nanocrystals in the diameter range of 0.2-1.5 nm, in order to identify the best conditions for optical emission. We show that the recombination rates of hydrogenated nanocrystals follow the quantum confinement feature in which the nanocrystal diameter is the principal quantity in determining the system response. Interestingly, a completely different behavior emerges from the OH-terminated or SiO$_2$-embedded nanocrystals, where the number of oxygens at the interface seems intimately connected to the recombination rates, resulting the most important quantity for the characterization of the optical yield in such systems. Besides, additional conditions for the achievement of high rates are constituted by a high crystallinity of the nanocrystals and by high confinement energies (small diameters).
\end{abstract}
\maketitle

\begin{table*}[t!]
  \centering
    \begin{tabular}[t]{c @{\hspace{0.6cm}} c @{\hspace{0.3cm}} c @{\hspace{0.3cm}} c @{\hspace{0.3cm}} c @{\hspace{0.3cm}} c @{\hspace{0.3cm}} c @{\hspace{0.3cm}} c}
      \toprule
        Structure         & NC silicons & NC-core silicons & Si-centered & interface-O & bridge-bonds & $d$ (nm) & $V_s$ (nm$^3$)\\
      \midrule
	Si$_{2}$/SiO$_2$      & 2       & 0                & No          & 6           & 0            & 0.2      & 3.26\\
	Si$_{5}$/SiO$_2$      & 5       & 1                & Yes         & 12          & 0            & 0.4      & 3.20\\
	Si$_{10}$/SiO$_2$     & 10      & 0                & No          & 16          & 0            & 0.6      & 2.65\\
	Si$_{17}$/SiO$_2$     & 17      & 5                & Yes         & 36          & 0            & 0.8      & 2.61\\
	Si$_{32}$/SiO$_2$     & 32      & 12               & No          & 56          & 0            & 1.0      & 8.72\\
	Si$_{35}$/SiO$_2$     & 35      & 5                & Yes         & 36          & 0            & 1.0      & 8.52\\
      \midrule
	a-Si$_{2}$/a-SiO$_2$  & 2       & 0                & No          & 6           & 0            & 0.2      & 2.74\\
	a-Si$_{5}$/a-SiO$_2$  & 5       & 1                & Yes         & 12          & 0            & 0.4      & 2.68\\
	a-Si$_{10}$/a-SiO$_2$ & 10      & 1                & Yes         & 20          & 0            & 0.6      & 2.61\\
	a-Si$_{17}$/a-SiO$_2$ & 17      & 5                & Yes         & 33          & 3            & 0.8      & 2.49\\
	a-Si$_{32}$/a-SiO$_2$ & 32      & 7                & No          & 45          & 3            & 1.0      & 8.67\\
      \midrule
	Si$_{10}$             & 10      & 0                & No          & 16          & 0            & 0.6      & 2.95\\
	Si$_{17}$             & 17      & 5                & Yes         & 36          & 0            & 0.8      & 9.48\\
	Si$_{26}$             & 26      & 10               & No          & 48          & 0            & 0.9      & 9.48\\
	Si$_{29}$             & 29      & 5                & Yes         & 36          & 0            & 1.0      & 9.48\\
	Si$_{32}$             & 32      & 12               & No          & 56          & 0            & 1.0      & 9.48\\
	Si$_{35}$             & 35      & 5                & Yes         & 36          & 0            & 1.0      & 9.48\\
	Si$_{47}$             & 47      & 17               & Yes         & 60          & 0            & 1.2      & 18.5\\
	Si$_{71}$             & 71      & 29               & Yes         & 84          & 0            & 1.4      & 32.0\\
	Si$_{87}$             & 87      & 35               & Yes         & 76          & 0            & 1.5      & 32.0\\
      \bottomrule
    \end{tabular}
    \caption{\it Structural characteristics for the crystalline embedded NCs (top set), amorphous embedded NCs (center set), crystalline suspended NCs (bottom set). For each structure are reported, respectively: number of NC silicons, number of core silicons (not bonded with oxygens), centered or not on one silicon, number of oxygens bonded to the NC, number of oxygens bridging two NC silicons, average diameter $d$, supercell volume $V_s$.} \label{table1}
\end{table*}

\section{Introduction}
The indirect nature of the energy band-gap in silicon has always been the major obstacle for its employment in optical devices, since the momentum conservation requires additional mechanisms involved in the recombination process (e.g. electron-phonon interaction), that occur with low probability, hence producing very poor emitting rates. In the last decade the discovery of efficient visible photoluminescence (PL) and optical gain from silicon nanocrystals (NCs) have demonstrated the possibility to overcome the indirect band gap of silicon by exploiting the quantistic behaviour of the matter at the nanoscale.\cite{pavesi, pavlockwood} Theoretically, the optical emission has been attributed to transitions between states localized inside the nanocrystal [as a consequence of the so-called quantum confinement (QC) effect],\cite{klimov, moskalenko, hill_whaley, derr_rosei} or between defect states.\cite{dovrat, kanemitsu, iwayama, averboukh, koponen, borczyskowski, godefroo} While there is still some debate on which of the above mechanisms primarily determines the emission energy, some works have proposed that a concomitance of both mechanisms is always present, favouring one or the other depending on the structural conditions.\cite{wolkin, allan, hao_green, gourbilleau, zhou, lin_chen, rolver}
\\Embedding Si-NCs in wide band-gap insulators is one way to obtain a strong QC. Si-NCs embedded in a silica matrix have been obtained with several techniques as ion implantation,\cite{brongersma,iwayama} chemical vapour deposition,\cite{hernandez,rolver,godefroo,gardelis} laser pyrolysis,\cite{delerue,kanemitsu} electron beam lithography,\cite{sychugov_linnros} sputtering,\cite{averboukh,antonova} and some others. Independently on the fabrication technique, in experimental samples no two NCs are the same, practically binding the measures to collective properties.
Experimentally, several factors contribute to make the interpretation of measurements on these systems a difficult task. For instance, samples show a strong dispersion in the NC size, that is difficult to be determined. In this case it is possible that the observed quantity does not correspond exactly to the mean size but instead to the most responsive NCs.\cite{credo} Again, NCs synthesized by using different techniques often show different properties in size, shape and in the interface structure. Also, in solid nanocrystal arrays some collective effects caused by electron, photon, and phonon transfer between the dots can strongly influence the luminescence dynamics in comparison with the case of isolated NCs.\cite{iwayama}
\\Recently, some studies on the PL spectra of individual silicon quantum dots have reported linewidths of 2 meV at $T$$\simeq$35 K, clearly below $k_{\text{\tiny{B}}}T$ at this temperature, demonstrating true quantum dot PL emission characteristics.\cite{delerue, sychugov_linnros, empedocles_norris_bawendi} The treatment of single NCs with atomic-like models allows a comparison between the collective response of a sample with the sum of the responses of the individual NCs, leading to useful informations about the NC-NC interaction mechanisms. Also, many works have already indicated the predominant contribution from Si-NC with particular size/shape,\cite{credo,reipa,nayfeh2} providing important implications for the identification of the best conditions for light emission and optical gain.\\

\noindent In this work we calculate the recombination-rates (RRs) of several Si-NCs in order to provide a description of the optical emission yield of such systems. Then, a systematic investigation on the different conditions of size, oxidation, strain, and amorphization is performed, allowing the identification of the circumstances that maximize the luminescence.
In Sec. \ref{sec_method} the structures and the methods used to calculate the emission rates are described. The results are presented and discussed in Sec. \ref{sec_results}. Conclusions are reported in Sec. \ref{sec_conclusions}.

\section{Structures and Method}\label{sec_method}
We built three different sets of NCs (see Table \ref{table1}): crystalline NCs embedded in SiO$_2$ (generated from betacristobalite), amorphous NCs embedded in silica (generated from a glass), OH-terminated NCs suspended in vacuum. A further set of hydrogenated NCs is obtained from the latter by replacing the OH groups with hydrogens and a subsequent geometric relaxation.
The crystalline embedded structures have been obtained from a betacristobalite cubic matrix by removing all the oxygens included in a cutoff-sphere, whose radius determines the size of the NC. By centering the cutoff-sphere on one silicon or in an interstitial position it is possible to obtain structures with different symmetries. To guarantee a proper shielding of the introduced strain by the surrounding silica we preserved a separation of about 1 nm between the NCs replica.
\\The glass models have been generated using classical molecular dynamics simulations of quenching from a melt, followed by \textit{ab-initio} relaxations (See Ref. \onlinecite{PRB1} for further details). The amorphous dot structures have been obtained, as for the crystalline systems, by applying the cutoff sphere on the glass supercells. It is worth to note that also in the crystalline case the embedding matrix gets amorphized after the inclusion of the NC, due to the metastable nature of the betacristobalite.\cite{PRB1}
\\The hydroxided NCs have been obtained by cutting off a spherical region from a bulk-silicon supercell, and successively passivating the dangling bonds with OH groups.
\\The relaxation of all the structures have been performed using the SIESTA code\cite{siesta1,siesta2} and Troullier-Martins pseudopotentials with non-linear core corrections. A cutoff of 150 Ry on the density and no additional external pressure or stress were applied. Atomic positions and cell parameters have been left totally free to move.\\
\\Each set is formed by NCs of different size, interface configuration, and symmetry. While the variation of the size is straightforwardly connected to the confinement energy, the latter aspects are associated to the formation of defect-states and to the localization of the electron-hole pairs.
\\For each structure we calculated the eigenvalues and eigenfunctions of the independent-particle Hamiltonian, using the Density Functional Theory (DFT), with the ESPRESSO package.\cite{espresso} Calculations have been performed using norm-conserving pseudopotentials within the Local Density Approximation (LDA) with a Ceperley-Alder exchange-correlation potential, as parametrized by Perdew-Zunger. An energy cutoff of 60 Ry on the plane wave basis have been considered.
\\It is well known that the DFT-LDA severely underestimates the band gaps for semiconductors and insulators: since DFT is a ground state theory, the electronic energy eigenvalues do not correctly describe the excitation energies of electron and holes.
A correction to the fundamental band gap is usually obtained by calculating the separate electron and hole quasiparticle energies via the GW method.\cite{review_onida_reining_rubio} The true quasiparticle energies, however, are still not sufficient to correctly describe a process in which electron-hole pairs are created. Their interaction can lead to a dramatic shift of peak positions as well as to distortions of the spectral lineshape.
Besides, previous many-body calculations on Si-NCs reported absorption spectra very close to the DFT-RPA calculated ones.\cite{delerue_lannoo_allan, ramos_paier_kresse_bechstedt, PRB1} In Ref. \onlinecite{PRB1} we verified this statement for the Si$_{10}$ and a-Si$_{10}$ embedded NCs, showing that the self-energy (calculated through the GW method) and electron-hole Coulomb (calculated through the Bethe-Salpeter equation) corrections exactly cancel out each other (with a total correction to the gap smaller than 0.1 eV) when the local field effects (LFE) are neglected. Furthermore, some our recent calculations (still unpublished), show that the LFE actually blue-shifts the absorption spectrum of the smallest systems (d$<$1 nm), with corrections of the order of few tenths of eV. Instead, for larger NCs no blue-shift is observed. Therefore, while such correction should be taken into account for a rigorous calculation of the RRs, we expect that the LFE would affect the final rates by only a few percent. Moreover, an enlargement of the gap for the smallest NCs would increase the calculated RRs, strengthening our conclusions about the size effects.
\\The reasons discussed above justify the choice of DFT-LDA for our calculations, allowing a good compromise between results accuracy and computational effort.\\

\noindent For a first-order radiative process, the spontaneous decay probability for unit time from a state $|i\rangle$ to a state $|j\rangle$, with $E_{ij}=E_i-E_j>0$, is defined from the Fermi golden rule and is given by\cite{dexter}
\begin{equation}
A_{ij} = \frac{16 \pi^2 n e^2 E_{ij}}{3 m^2 h^2 c^3}~ |\langle j | \textbf{p} | i \rangle |^2 ~, \label{eq.spont_prob}
\end{equation}
where $n$,$e$,$m$,$h$,$c$,$\textbf{p}$ are the refractive index, electron charge, electron mass, Planck constant, light speed, and momentum operator, respectively. The static refractive index has been obtained from the real part of the dielectric function at zero energy, $n=\sqrt{\varepsilon_1(\omega=0)}$, calculated in the random-phase approximation (RPA). We observed an increasing trend of $n$ with the NC diameter, with $1.1<n<1.2$ ($1.75<n<1.85$) for the suspended (embedded) NC sets. Considering the large variation of the final emission rates (practically determined by the matrix elements) we approximated $n=1$ for all the calculations. Equation \ref{eq.spont_prob} does not take into account any summation over the k-points of the reciprocal space, since the states belonging to the NC are strogly localized (due to the high QC), and no energy dispersion in the bands occurs (see Ref. \onlinecite{PRB1}). For this reason we performed the calculation of the probabilites at the single $\Gamma$ point.
\\From Eq. \ref{eq.spont_prob} we can define the thermally averaged recombination rate by \cite{delerue_allan_lannoo}:
\begin{equation}
\left\langle\frac{1}{\tau}\right\rangle=\frac{\sum_{i,j}A_{ij}\exp\left(-\frac{E_{ij}}{k_{\text{\tiny{B}}}T}\right)}{\sum_{i,j}\exp\left(-\frac{E_{ij}}{k_{\text{\tiny{B}}}T}\right)} ~, \label{eq.recombination_rate}
\end{equation}
where $i$ and $j$ are states belonging to the conduction and valence bands, respectively. The spin is not included in the RR calculation, and therefore the effects due to exchange interaction are neglected, with a possible influence on the low temperature ($<50$ K) dependence of the calculated rates.\cite{calcott}
Morerover, while the recombination through nonradiative decay mechanisms is expected to be favoured for high temperatures, all experimental results indicate that radiative transitions dominate for a broad temperature range, up to room temperature.\cite{saar,driel}
Finally, Eq. \ref{eq.recombination_rate} assumes that the thermalization of the electron and the hole after the excitation in the bands is more efficient than the radiative recombination. This condition is reasonably satisfied because the radiative recombination is usually not very efficient: RRs for spherical crystallites of different diameters have been reported ranging from about 0 to 0.1 ns$^{-1}$,~~\cite{delerue_allan_lannoo} compared to timescales of the order of 1 ps for thermally-activated processes.

\section{Results}\label{sec_results}

\begin{table}[b!]
  \centering
    \begin{tabular}[t]{c @{\hspace{0.6cm}} c @{\hspace{0.3cm}} c @{\hspace{0.3cm}} c @{\hspace{0.3cm}} c @{\hspace{0.3cm}} c}
      \toprule
        Structure & $\Omega$ & gap-H  & gap-OH \\
      \midrule
	Si$_{10}$ & 1.6      & 4.61    & 1.97 \\
	Si$_{17}$ & 3.0      & 4.06    & 3.50 \\
	Si$_{26}$ & 3.0      & 3.70    & 3.05 \\
	Si$_{29}$ & 1.5      & 3.44    & 1.89 \\
	Si$_{32}$ & 2.8      & 3.54    & 2.93 \\
	Si$_{35}$ & 1.2      & 3.43    & 1.50 \\
	Si$_{47}$ & 2.0      & 3.10    & 1.83 \\
	Si$_{71}$ & 3.0      & 2.76    & 2.26 \\
	Si$_{87}$ & 1.46     & 2.45    & 1.19 \\
      \bottomrule
    \end{tabular}
    \caption{\it Oxydation degree $\Omega$, and HOMO-LUMO gaps for the H- and OH-terminated NCs. }\label{table.fig.recombrates_suspended}
\end{table}

\begin{figure}[b!]
  \centerline{\includegraphics[draft=false,height=6.8cm,angle=270]{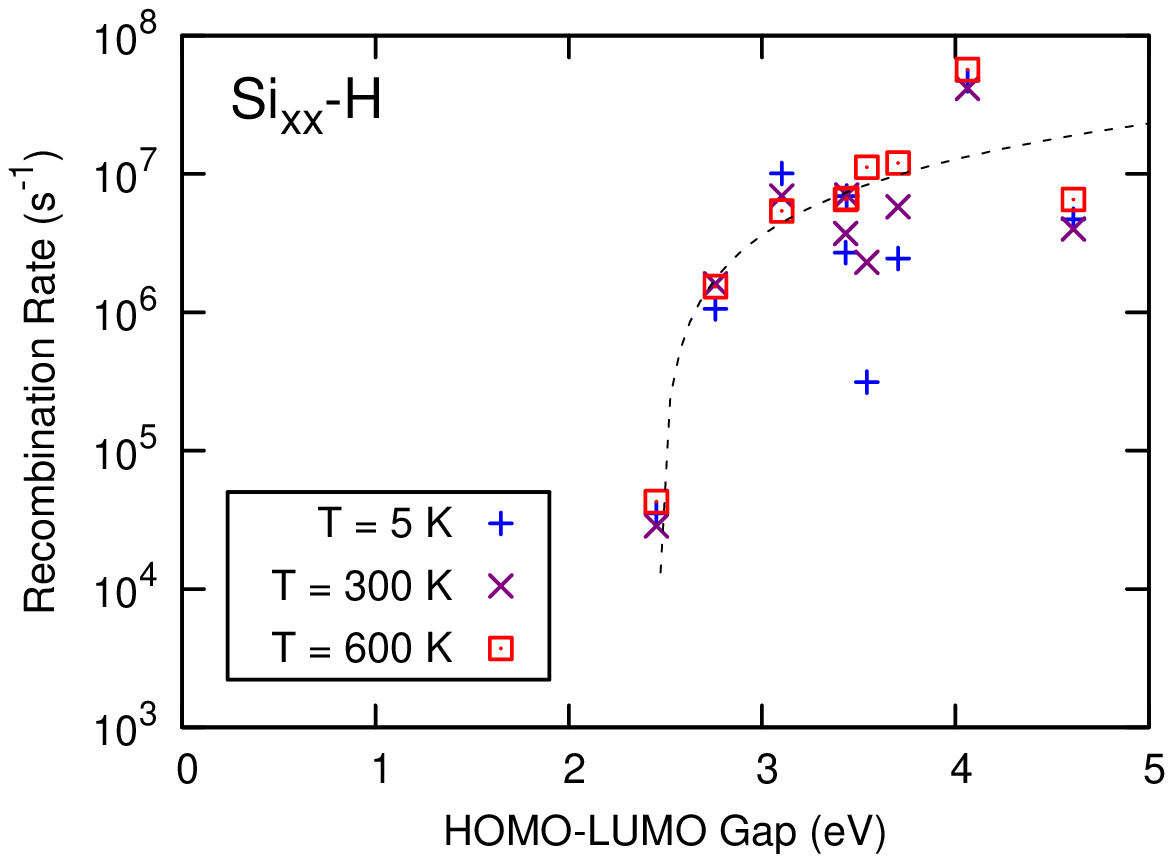}{\bf a)}}
  \centerline{\includegraphics[draft=false,height=6.8cm,angle=270]{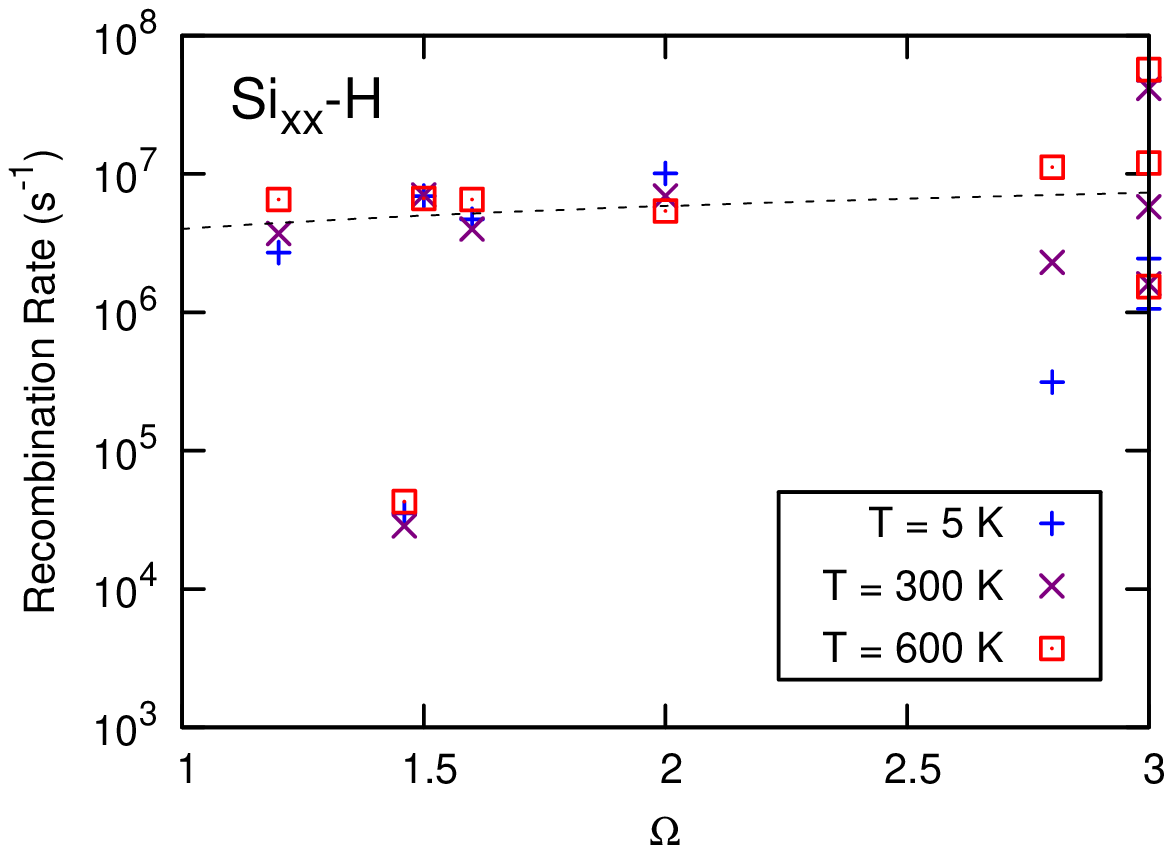}{\bf b)}}
  \centerline{\includegraphics[draft=false,height=6.8cm,angle=270]{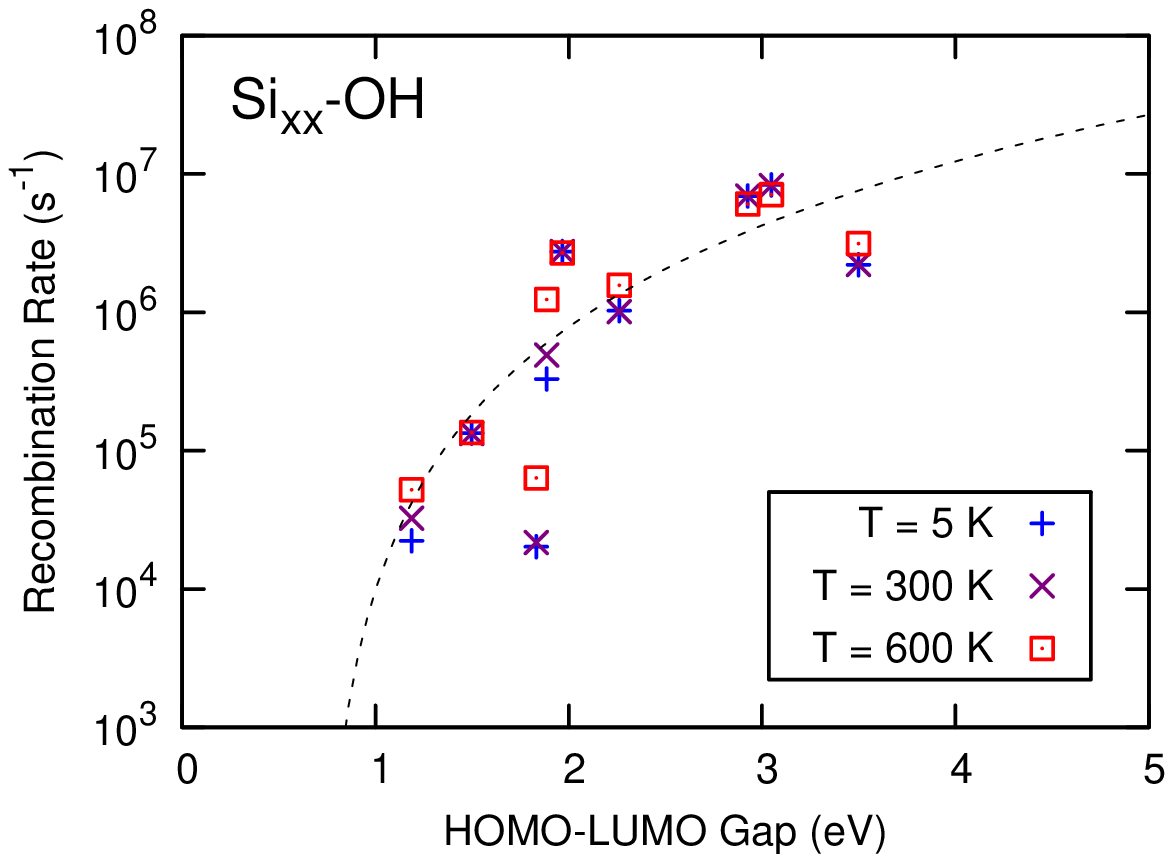}{\bf c)}}
  \centerline{\includegraphics[draft=false,height=6.8cm,angle=270]{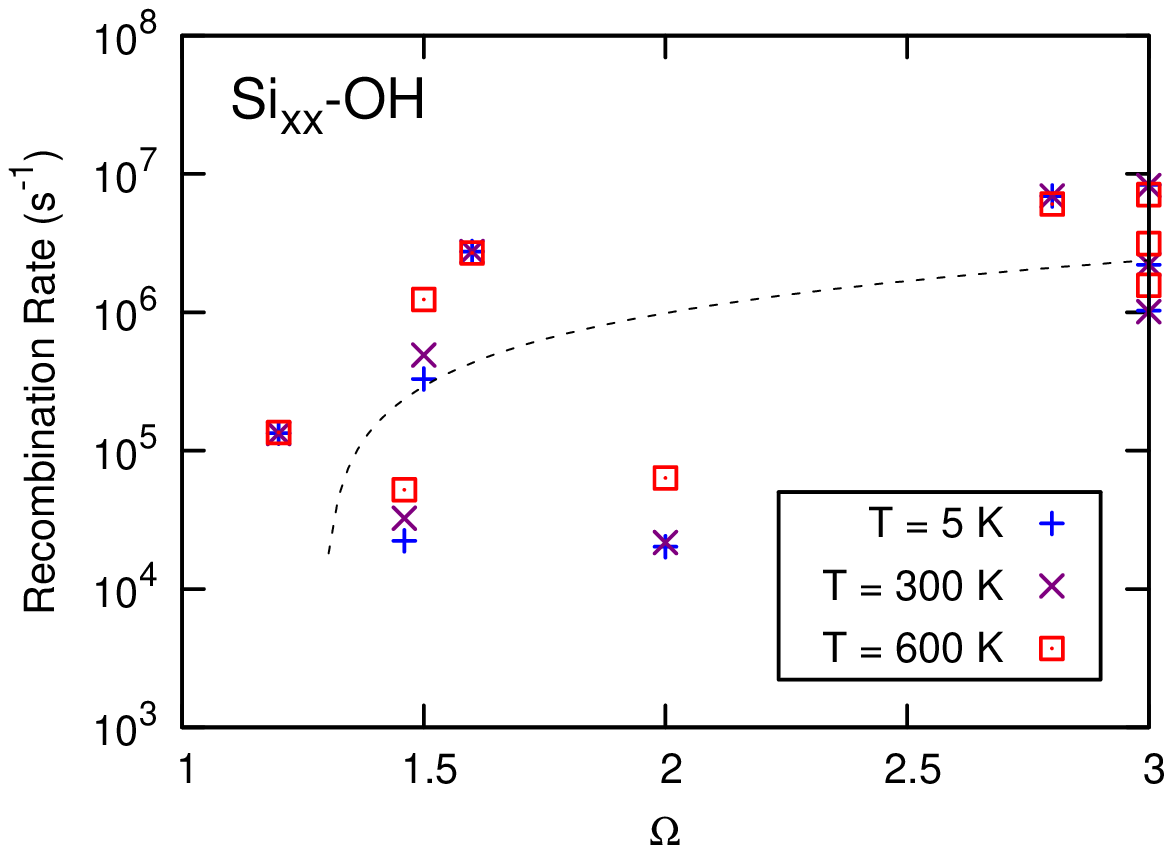}{\bf d)}}
  \caption{\it Thermally averaged recombination rates, calculated at different temperatures, for H-terminated [a) and b)] and OH-terminated [c) and d)] NCs, as a function of the HOMO-LUMO energy gap or of the oxidation/hydrogenation degree, $\Omega$. The dashed line stands for the power-law fit to the data at $T=300$ K (see text).
  }\label{fig.recombrates_suspended}
\end{figure}

In this Section we present the thermally averaged RRs, calculated through Eq. \ref{eq.recombination_rate}, for the whole set of NCs. Figure \ref{fig.recombrates_suspended} reports the RRs at the temperatures of 5 K, 300 K, and 600 K, as a function of the energy gap, $E_G$ [given by the difference between the LUMO (Lowest Unoccupied Molecular Orbital) and the HOMO (Highest Occupied Molecular Orbital) states], and of the oxidation (or hydrogenation) degree, $\Omega$ [given by the ratio between the oxygens (or hydrogens) at the interface by the number of silicons to which they are connected], for the OH-terminated and H-terminated NCs. The $E_G$ and $\Omega$ values for each suspended NC are reported in Table \ref{table.fig.recombrates_suspended}. Some trends clearly emerge from the rates reported in Fig. \ref{fig.recombrates_suspended}.
The simplest case of hydrogenated NCs shows a clear increase of the emission rate with the gap (Fig. \ref{fig.recombrates_suspended}a), indicating that the smaller NCs are the more efficient, with characteristic times even below 0.1 $\mu$s. 
A power-law fit to the data at $T=300$ K results $7.7\cdot10^6(E_G/$eV$-2.47)^{1.19}$ s$^{-1}$ (Fig. \ref{fig.recombrates_suspended}a, dashed line).
By considering the proportionality introduced by the $E_{ij}$ term in Eq. \ref{eq.spont_prob}, the fit points out an almost constant matrix element with the NC size, as occurring for an ideal two-level exciton system.\cite{driel} Besides the general trend, some scattering is present in the data, that can be better interpreted by examining the RR-vs-$\Omega$ dependence (Fig. \ref{fig.recombrates_suspended}b). It is well known that the HOMO-LUMO gap of hydrogenated NCs is not related to the passivation regime, simply following the QC.\cite{PRB2} This uncorrelation is well reflected in the RR-vs-$\Omega$ plot, showing a much clearer picture in that most of the NCs emits at the same rate (of about $10^7$ s$^{-1}$) independently of $\Omega$. Anyway, for NCs with the same $\Omega$ the characteristic time increases with size, as shown for $\Omega$$\simeq$1.5 and $\Omega$$=$3.
We note that the scattering of the data is more pronounced at low temperatures, where the RR is mostly due to the LUMO-HOMO recombination. Such behavior is due to the fact that the lowest transition corresponds rarely to the most efficient among all the possible transitions. Contrary, at high temperatures more channels contribute to the average, leading to a smoother trend of the RR. Quantitatively, the variation of the RR in the temperature range of 300-600 K is always smaller than one order of magnitude, making such systems poorly dependent on $T$. The above results for H-terminated NCs are in good agreement with calculations on porous silicon in the tight-binding approximation,\cite{delerue_allan_lannoo} and with other DFT-based calculations on Si/Ge NCs.\cite{weissker3}
\\The analysis of the data relative to the hydroxided NCs follows that of the hydrogenated ones. The RR dependence on $E_G$ (Fig. \ref{fig.recombrates_suspended}c) shows a supra-linear trend, with some noticeable deviations induced by the $E_G$-$\Omega$ correlation. The power-law fit to the data at $T=300$ K results~ $3.7\cdot10^5(E_G/$eV$-0.71)^{2.94}$ s$^{-1}$, very near to the cubic relation occurring for an ideal two-level atom.\cite{driel} Besides the QC effect, clearly dominating for hydrogenated NCs, in the case of hydroxided NCs we have to consider that the band-gap is strongly correlated to the oxidation degree, producing strong fluctuations of the gap with the NC size.\cite{PRB2} Such correlation could be responsible for the large broadening of the PL spectra that has been observed experimentally, also at low temperatures,\cite{brongersma, hernandez, rolver, gardelis, delerue, averboukh, dovrat, klimov, lin_chen, kanemitsu2, ray_green, nayfeh} even for apparently monodispersed multilayered samples.\cite{cazzanelli} This consideration reflects on the RR-vs-$\Omega$ dependence (Fig. \ref{fig.recombrates_suspended}d). Contrary to the hydrogenated case, here some dependency on $\Omega$ of the RR occurs, with some correlation emerging between the two quantities. In summary, we observe that the hydrogenated NCs present higher rates with respect to OH-terminated NCs, as experimentally demonstrated.\cite{nayfeh,dohnalova} Besides, for the hydroxided NCs the optical emission is favoured when a high O/Si ratio is achieved, in well agreement with some experimental observations.\cite{khriachtchev2,hao_green}\\

\begin{figure}[b!]
  {\bf a)}\includegraphics[draft=false,width=5.5cm,angle=270]{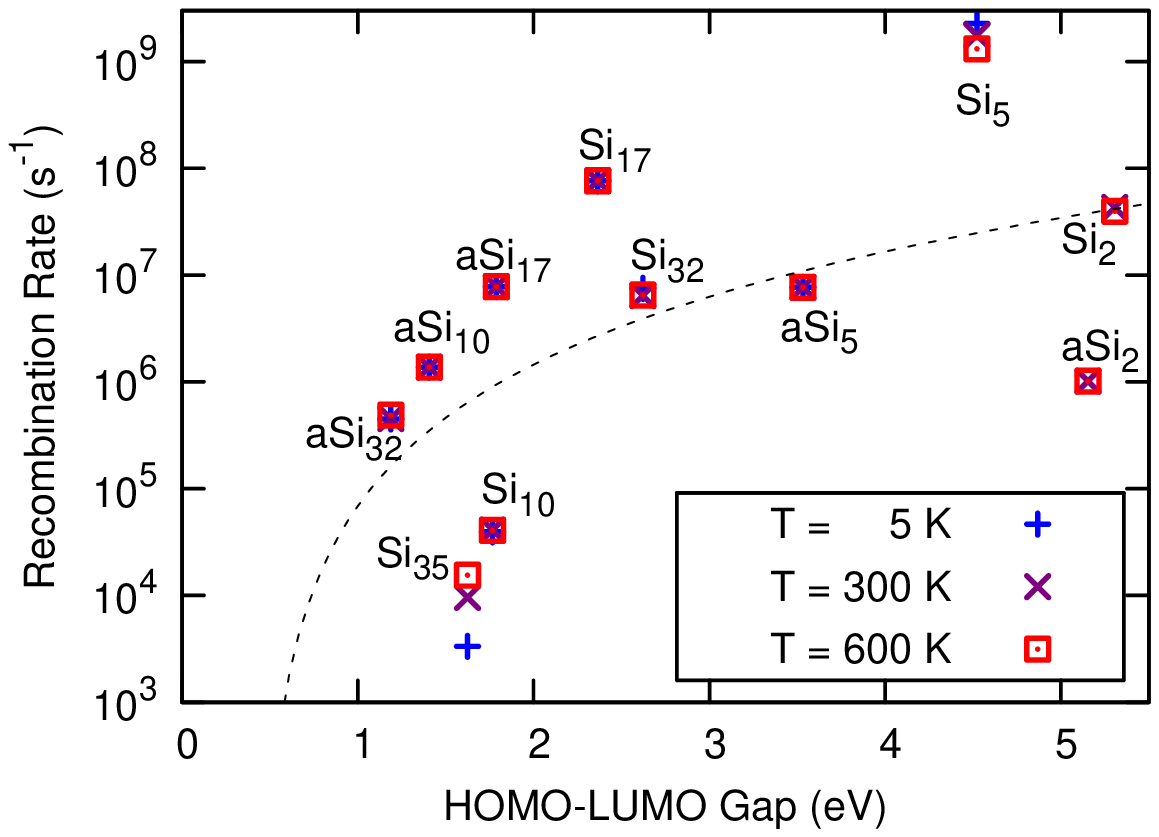}\\
  {\bf b)}\includegraphics[draft=false,width=5.5cm,angle=270]{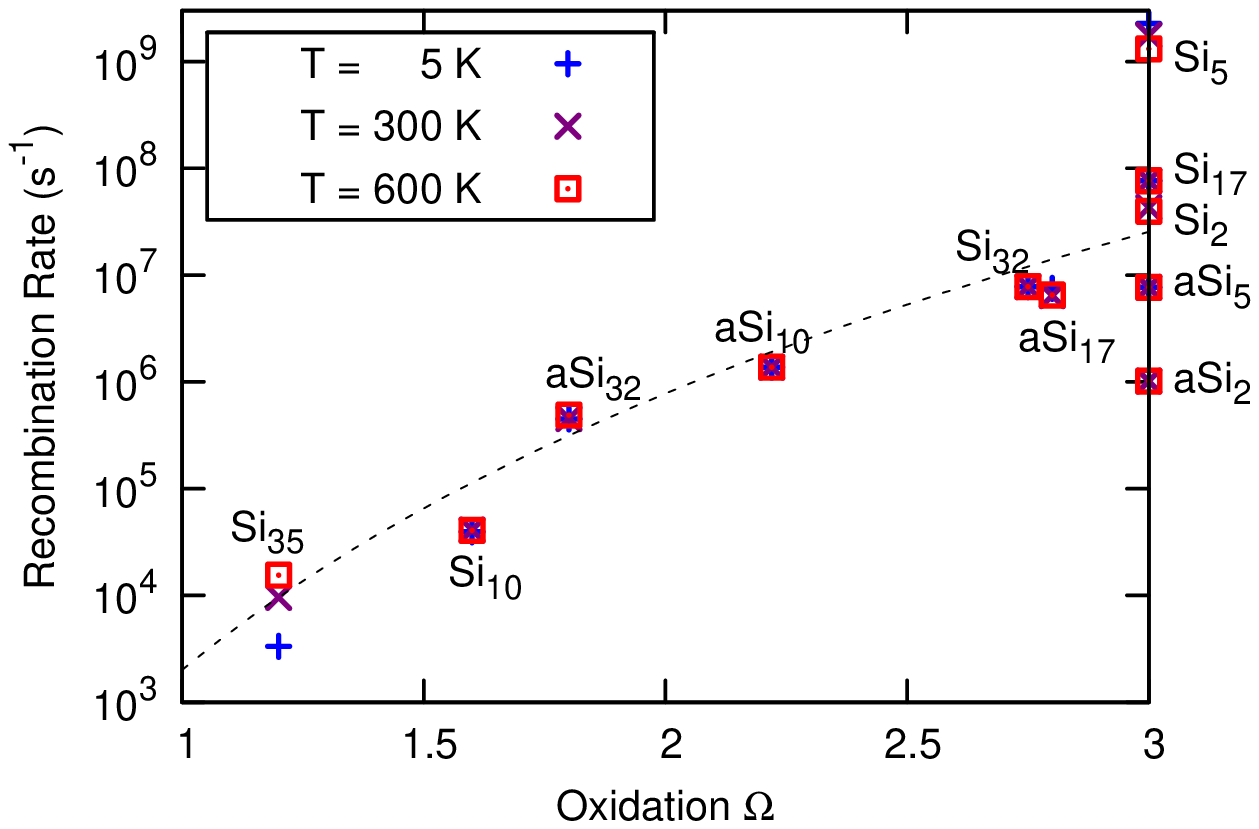}
  \caption{\it Thermally averaged recombination rates, calculated at different temperatures, for the Si/SiO$_2$ embedded NCs, a) as a function of the HOMO-LUMO energy gap, and b) as a function of the oxidation degree, $\Omega$. The dashed line stands for the power-law fit to the data at $T=300$ K (see text). }\label{fig.recombrates_embedded}
\end{figure}

\noindent The calculation of the RRs for the embedded structures, in comparison with those of the hydroxided-suspended NCs, could shed some light on the role of the embedding matrix. For example, it is known that the strain induced by the silica on the NC tends to red-shift the band-gap of the smallest NCs ($<1$ nm) while blue-shifting that of the largest ones ($>2$ nm).\cite{PRB2,peng,franceschetti} Therefore, the effects of such strain on the emission properties are investigated by comparing the results for the suspended (unstrained) and the embedded systems. Figure \ref{fig.recombrates_embedded} reports the RRs for the embedded NCs set, at the temperatures of 5 K, 300 K, and 600 K, as a function of the HOMO-LUMO gap (top panel), and of $\Omega$ (bottom panel). In this case $\Omega$ is also an index for the strain induced on the NC by the silica matrix, as discussed in Ref. \onlinecite{PRB2} where a correlation between the two has been proposed. The calculations for the embedded structures confirm an increase of the rates with $E_G$ and $\Omega$. The fit to the data at $T=300$ K results~ $4.0\cdot10^5(E_G/$eV$-0.45)^{2.94}$ s$^{-1}$, very close to what obtained for the hydroxided NCs. Therefore, the strain seems to not modify the picture, discussed above, regarding the effects of size and oxidation. The two smallest, highly oxidized NCs (Si$_2$ and Si$_5$) follows the trend, attaining impressive characteristic times, even below 1 ns (Si$_5$ case). Nevertheless, the embedding matrix has some effects on the final rates, that are sometimes strongly increased (Si$_{17}$), sometimes unaffected (Si$_{32}$), and other times strongly reduced (Si$_{10}$ and Si$_{35}$). From the comparison of Fig. \ref{fig.recombrates_suspended}d and Fig. \ref{fig.recombrates_embedded}b, we note that the introduction of the matrix produces a clear RR-vs-$\Omega$ trend, revealing that the oxidation is a crucial quantity for the characterization of the optical properties of Si/SiO$_2$ systems.
\\The opportunity of considering two opposite ideal cases of cristalline and amorphous NCs comes from the fact that real samples are characterized by a certain amount of amorphization, especially for smaller NCs.\cite{rinnert,gourbilleau,veprek} Important connections of the amorphization with the PL spectrum have been reported, generally indicating, for amorphous NCs, reduced\cite{meldrum,anthony} or comparable\cite{kanemitsu2,rinnert,dhara} PL intensities with respect to the cristalline case.
\\Our results show that the introduction of disorder seems to flatten the RR trend, with all the values oscillating within 10$^6$-10$^7$ s$^{-1}$. We note that the higher emission of the a-Si$_{10}$ NC with respect to the Si$_{10}$ case can be explained with the larger oxidation degree of the amorphized NCs (see bottom panel of Fig. \ref{fig.recombrates_embedded}). The same rule applies to the a-Si$_{32}$ NC, presenting a smaller RR due to a reduced oxidation with respect to its crystalline counterpart. Instead, when the oxidation degree is preserved, the disorder always reduces the RR (Si$_2$, Si$_5$). Therefore, our results confirms the experimental observations, indicating that the amorphization process generally reduces the optical yield.

\section{Conclusions}\label{sec_conclusions}

We evaluated the recombination rates at different temperatures for Si-NCs of diameters in the range of 0.2-1.5 nm, as for H- and OH-terminated NCs, like as for NCs embedded in a SiO$_2$ matrix. Different conditions of size, passivation, strain, and symmetry have been considered in order to investigate for the best conditions of radiative emission.
\\We found that the rates follow a trend with the emission energy that is nearly linear for the hydrogenated NCs, and nearly cubic for the NCs passivated with OH groups or embedded in SiO$_2$. Moreover, the hydrogenic passivation produces higher optical yields with respect to the hydroxilic one, as evidenced experimentally.\cite{nayfeh,dohnalova} Besides, for the hydroxided NCs the emission is favoured for systems with a high O/Si ratio.\cite{khriachtchev2,hao_green}
\\The analysis of the results for the embedded NCs reveals a clear picture in which the smallest, highly oxidized, crystalline NCs, belong to the class of the most optically-active Si/SiO$_2$ structures, attaining impressive rates of less than 1 ns, in nice agreement with experimental observations.\cite{meldrum,anthony,kanemitsu2,rinnert,dhara}

\section*{ACKNOWLEDGEMENTS}
We thank Ivan Marri, Elena Degoli, and Marco Govoni for a careful reading of the manuscript. R.G. acknowledge financial support from Fondazione Cassa di Risparmio di Modena under the project "Progettazione di materiali nanostrutturati semiconduttori per la fotonica, l'energia rinnovabile e l'ambiente". This work is supported by PRIN2007 and by Ministero degli Affari Esteri, Direzione Generale per la Promozione e la Cooperazione Culturale.

\end{document}